\documentclass[prl,twocolumn,showpacs,tightenlines]{revtex4}

\usepackage{amsfonts}
\usepackage{amsmath}
\usepackage{amssymb}
\usepackage{graphicx}

\DeclareSymbolFont{rsfs}{U}{rsfs}{m}{n}
\DeclareSymbolFontAlphabet{\mathcal}{rsfs}

\begin{document}
\title[Creation of maximally entangled states.]
{Controlled creation of maximally entangled states using a SQUID ring coupled
  to an electromagnetic field}
\author{M.J. Everitt}
\email{m.j.everitt@sussex.ac.uk}
\author{T.D. Clark}
\email{t.d.clark@sussex.ac.uk}
\author{P.B. Stiffell}
\author{H. Prance}
\author{R.J. Prance}
\affiliation{\mbox{Quantum Circuits Group, School of Engineering, University of Sussex, Falmer, Brighton, BN1 9QT, U.K.}}
\author{J.F. Ralph}
\affiliation{\mbox{Department of Electrical and Electronic Engineering, Liverpool University, Brownlow Hill, Liverpool L69 3GJ, U.K.}}

\begin{abstract}
  We present solutions  of the time dependent Schr\"{o}dinger equation
  for  a SQUID ring coupled  to an electromagnetic field, both treated
  quantum mechanically.  We that  show the SQUID  ring can  be used to
  create a  maximally entangled state with  the em field  that without
  dissipation remains constant    in time. Using methods familiar   in
  quantum optics,   we extend the    model to include  the  effects of
  coupling this  system to a dissipative  environment. With this model
  we   show that although   such  an  environment  makes  a noticeable
  difference to the time evolution of the system,  it need not destroy
  the the  entanglement   of this   coupled system  over   time scales
  required for quantum technologies.
\end{abstract}
\maketitle

Experiments   to  probe  the  quantum  properties  of  superconducting
Josephson                 weak           link                  (SQUID)
circuits~\cite{FriedmanPCTL00,vanderWalWSHOLM00,RouseHL95,SilvestriniRGE00,NakamuraPT99}
(in this work  a superconducting ring   enclosing a single weak  link)
have highlighted the possible use of such  devices in emerging quantum
technologies  such   as     quantum  computation,  communication   and
information  processing~\cite{OrlandoMTvLLM99,MakhlinSS99,AverinNO90}. 
In an earlier  paper, motivated in   part by experiment  we solved the
time dependent Schr\"{o}dinger equation  (TDSE) for an isolated  SQUID
ring    modulated   by    a    classical,  monochromatic,    microwave
field~\cite{ClarkDREPPWS98}.                                    Recent
studies~\cite{Everitt2001b,Al-SaidiS02}  of a fully quantum mechanical
model of electromagnetic (em) fields coupled to SQUID rings have, from
a theoretical standpoint, furthered our understanding of the manner in
which such circuit may be utilised.

An essential feature  revealed   by these studies  is  that  a quantum
mechanical SQUID ring is highly non-perturbative in nature, a property
that  can be used  to  affect very  strongly the non-linear  dynamical
behaviour  of SQUID  ring based quantum  circuit systems. Furthermore,
these  effects can be  controlled through the external (bias) magnetic
flux  $\Phi_{x}$  applied to  the  ring.   Even so, although   several
interesting new results came from the investigation of coupled ring-em
field systems, in each case we chose, for ease of computation, a SQUID
ring Hamiltonian  description   which was time   independent. However,
there are many  circumstances   where  time  dependence needs  to   be
introduced into this description.   In  the work described  here  this
arises because we wish to entangle a SQUID ring with an em field mode.
As we will  show, this  can be  achieved by applying  a time dependent
magnetic flux   to  the SQUID   ring  which changes slowly   enough to
maintain the ring-field system in its current state but fast enough to
ensure strong  entanglement. In this paper  we solve the TDSE or, more
generally, the master equation  where dissipation involved, for tensor
products  of  SQUID ring-em   field  systems  with  a particular  time
dependent potential.   We show that we  can  use the   flux dependent,
non-perturbative  nature  of the   SQUID  ring  to  great utility   in
controlling its level  of entanglement with  an em field.  As we shall
also demonstrate, the  level of this entanglement  can decay with time
or  remain  constant  depending    on   whether or   not   dissipation
(decoherence) is present.

\begin{figure}[tb]
\begin{center}
\resizebox*{0.38\textwidth}{!}{\includegraphics{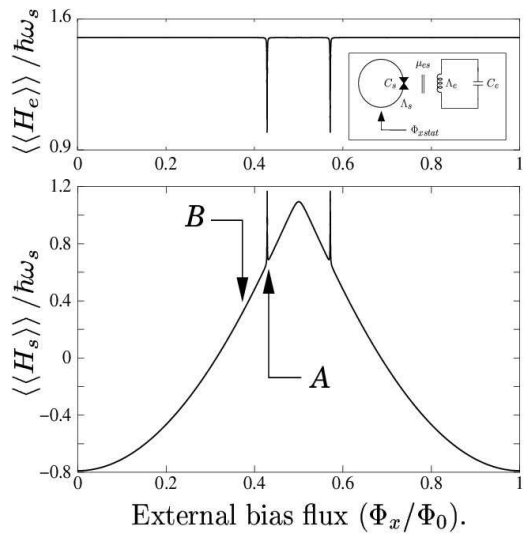}}
\caption{Time averaged energy expectation values for the initial state
$\left\vert 1_{e}0_{s}\right\rangle $ for (a) the em field and (b) the SQUID
ring as functions of $\Phi_{x}$.\vspace*{-0.3in}}
\label{fig:energy}
\end{center}
\end{figure}

In  dealing  with a SQUID ring   coupled to  an   em field  as quantum
mechanical objects we represent the field as  an equivalent circuit, a
parallel, inductor-capacitor (LC),  oscillator~\cite{Everitt2001b}. We
also assume that the coupling between the SQUID  ring and the em field
is inductive~\cite{Likharev}. The  schematic for  this coupled circuit
system is shown inset in~\ref{fig:energy}.

With reference to  the inset in  figure~\ref{fig:energy}, we can write
down a Hamiltonian  for the coupled  two mode (SQUID  ring  + em field
oscillator mode) system of the form
\begin{equation}
H=H_{e}+H_{s}-H_{es} \label{totalHamiltonian}
\end{equation}
where $H_{e}$ and $H_{s}$ are,  respectively, the Hamiltonians for the
em  field and the SQUID ring  and the  interaction energy between them
(for  our purposes purely  inductive) is $H_{es}$.  For these we adopt
the    convention  that operators  with   the   subscripts  $s$ or $e$
represent, respectively, those  quantities associated  with the  SQUID
ring or the em field.  We also note that  we use the first four energy
eigenstates of  each Hamiltonian to  represent  the individual circuit
component of the system.  This number  was  found to be  sufficient to
compute accurately the results presented in this paper.

The Hamiltonians for these quantum circuits  can be expressed in terms
of  the conjugate variables  of magnetic  flux$~\Phi$ (coordinate) and
electric                 displacement                    flux$~Q\left(
  \rightarrow-i\hbar\partial/\partial\Phi\right) $and  the appropriate
circuit parameters. Thus, for  a SQUID ring (inductance $\Lambda_{s}$,
weak  link capacitance  $C_{s}$) $\Phi$  is  the  total magnetic  flux
threading the ring  and $Q$ is  the  total electric displacement  flux
(displacement charge) between the electrodes of the weak link; for the
field oscillator these flux and charge variables relate, respectively,
to the magnetic  and   electric  displacement  fluxes  threading   the
equivalent   inductor  (inductance    $\Lambda_{e}$)   and   capacitor
(capacitance $C_{e}$). In each  case these operators satisfy the usual
cannonical commutation   relation  $\left[ \Phi,Q\right] =i\hbar$.  We
assume that both the SQUID  ring and em field oscillators (frequencies
$\omega_{s}/2\pi\left( =1/2\pi\sqrt{\Lambda_{s} C_{s}}\right)  $   and
$\omega_{e}/2\pi\left(   =1/2\pi\sqrt{\Lambda_{e}C_{e}    }\right)  $,
respectively)   operate  in the   quantum     regime, i.e.  such  that
$\hbar\omega_{s},\hbar\omega_{e}\gg k_{B}T$ for temperature $T$.

Given a SQUID  ring  inductance  of  $\Lambda_{s}$  and  a weak   link
capacitance   of $C_{s}$, the   SQUID ring Hamiltonian   is, as usual,
translated  in external bias  flux~\cite{ClarkDREPPWS98} and takes the
time dependent form
\begin{equation}
H_{s}=\frac{Q_{s}^{2}}{2C_{s}}+\frac{\Phi_{s}^{2}}{2\Lambda_{s}}-\hbar\nu
\cos\left[  \frac{2\pi}{\Phi_{0}}\left(  \Phi_{s}+\Phi_{x}(t)\right)  \right]
-Q_{s}\frac{\partial}{\partial t}\Phi_{x}(t). \label{squidHamiltonian}
\end{equation}
where   $\hbar\nu/2=0.0215~\Phi_{0}^{2}/\Lambda_{s}$ is the  Josephson
phase coherent coupling energy, $\Phi_{0}\left(  =h/2e\right) $ is the
superconducting flux  quantum and we note that  the last  term, $Q_{s}
\frac{\partial}{\partial t}\Phi_{x}(t)$  in   (\ref{squidHamiltonian})
arises  from our     translating  the  Hamiltonian  by   $\Phi_{x}(t)$
(see~\cite{ClarkDREPPWS98} for more details).  For this quantum regime
we  set  $C_{s}=1\times10^{-16}$F  and $\Lambda_{s}=3\times10^{-10}$H,
parameter values which  are appropriate for   a mesoscopic SQUID  ring
operating at a few K~\cite{Everitt2001b}.

The Hamiltonian for  the em field mode,  which is somewhat simpler but
equivalent to that for a simple harmonic oscillator, can be written as
\begin{equation}
H_{e}=\frac{Q_{e}^{2}}{2C_{e}}+\frac{\Phi_{e}^{2}}{2\Lambda_{e}}
\label{fieldHamiltonian}
\end{equation}
while  the energy associated with the   inductive coupling between the
ring and the em field mode is given by
\begin{equation}
H_{es}=\frac{\mu_{es}}{\Lambda_{s}}\Phi_{s}\Phi_{e} \label{couplingEnergy}
\end{equation}
where $\mu_{es}=0.01$ is the ring-field magnetic flux linkage factor. For
simplicity, and to facilitate the computations, we set $C_{e}=C_{s}$ and
$\Lambda_{e}=\Lambda_{s}$ throughout this work.

Using  (\ref{totalHamiltonian}),    we    showed   in   our   previous
work~\cite{Everitt2001b} that  energy could  be exchanged between  the
quantum modes of a SQUID ring-em field system around particular values
of the static flux applied to the ring. For the SQUID ring ($i=s$) and
the  em field ($i=e$) the  time averaged energy expectation values are
found from
\[
\left\langle \left\langle H_{i}\right\rangle \right\rangle =\lim
_{\tau\rightarrow\infty}\frac{1}{\tau}\int_{0}^{\tau}\left\langle \psi\left(
t\right)  \left\vert H_{i}\right\vert \psi\left(  t\right)  \right\rangle ~dt
\]
Examples of exchange (or transition)   regions in these time  averaged
energies are shown in figure~\ref{fig:energy} around the arrowed point
A (i.e.  where the energy difference between  two of the ring energies
is very  close to $n\hbar\omega_{e}$, $n$ integer).  As in the rest of
the  paper, the initial  state was taken  to be $\left\vert \psi\left(
    0\right) \right\rangle =\left\vert 1_{e}0_{s}\right\rangle $, i.e.
with the em field  mode in its first excited  Fock state and the SQUID
ring in its ground energy  state. In computing these averaged energies
we clearly could  not integrate for the system  over an infinite time;
instead we calculated the time averaged  expectation values for a time
$\tau$ such    that $\left\langle    \left\langle   H_{i}\right\rangle
\right\rangle $  did   not change noticeably   if $\tau$  was  further
increased.

The  actual value of $\Phi_{x}$ around  which exchange regions develop
depends  on  circuit (ring and field    oscillator) parameters and the
initial condition of the system.  For example, for the particular case
presented     in   figure~\ref{fig:energy}    the   point  $A$,     at
$\Phi_{x}=0.42864~\Phi_{0}$ (together     with    its   twin    at
$\Phi_{x}=0.57136\Phi_{0}$, as  shown), is where significant energy is
exchanged between the SQUID ring and the em field.

\begin{figure}[tb]
\begin{center}
\resizebox*{0.38\textwidth}{!}{\includegraphics{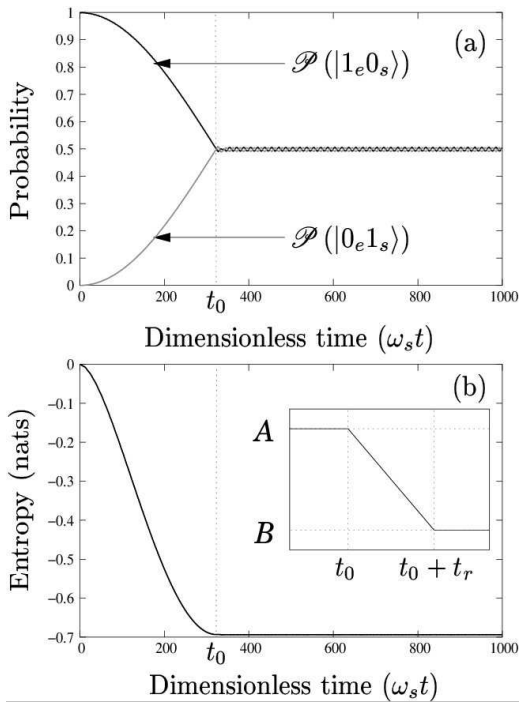}}
\caption{(a) Probability of the system being in state $\left\vert 1_{e}%
0_{s}\right\rangle $ or $\left\vert 0_{e}1_{s}\right\rangle $, i.e.
$\mathcal{P}\left\vert 1_{e}0_{s}\right\rangle $ or $\mathcal{P}\left\vert
0_{e}1_{s}\right\rangle $ (b) system entanglement entropies $I_{s}$ and
$I_{e}$ (here equal) evolving at point A for initial condition $\left\vert
1_{e}0_{s}\right\rangle $.\vspace*{-0.3in}}
\label{fig:probAndEntangle}
\end{center}
\end{figure}

The energy  exchange,  which reaches  a maximum at   the centre of the
exchange region, follows the  strength  of  the coupling between   the
SQUID ring and the em field mode. Other  subtle properties of this two
component quantum circuit system  also  follow this changing  coupling
strength, in particular the quantum  entanglement between the ring and
the field mode. Entanglement plays a central role in considerations of
quantum     technologies~\cite{OrlandoMTvLLM99,MakhlinSS99,AverinNO90}
where it is often of the first  importance that strong entanglement is
maintained as the quantum system evolves with time. With this in mind,
and as the central result in this paper, we set about trying to create
a state of  maximal entanglement that remains  constant with  time. In
our chosen system this takes the form
\begin{equation}
\left\vert \psi\left(  t\right)  \right\rangle =\frac{1}{\sqrt{2}}\left(
e^{i\varphi_{1}}\left\vert 1_{e}0_{s}\right\rangle +e^{i\varphi_{2}}\left\vert
0_{e}1_{s}\right\rangle \right)  \label{maxEntangledState}
\end{equation}
To implement this we now make the  external flux time dependent with a
functional form   (see inset  in figure~\ref{fig:probAndEntangle}~(b))
given by
\begin{equation}
\Phi_{x}\left(  t\right)  =\left\{
\begin{array}[c]{ll}
A & t\leq t_{0}\\
\frac{\left(  B-A\right)  }{t_{r}}\left(  t-t_{0}\right)  & t_{0}<t\leq
t_{0}+t_{r}\\
B & t>t_{0}+t_{r}
\end{array}
\right.  \label{fluxTimeFunction}
\end{equation}
where     $t_{0}=326\omega_{s}^{-1}$   and     the    ramp   time   is
$t_{r}=16.6\omega _{s}^{-1}$. Following this change  in bias flux, the
operating  point     has      moved from     $A$     to   $B$  $\left(
  {\Phi}_{x}=0.38\Phi_{0}\right)  $ in figure~\ref{fig:energy},
well outside the exchange region. The cut off time $t_{0}$ is the time
at which the  probability  of the  system being  in state  $\left\vert
  1_{e}0_{s}\right\rangle $ is equal to the probability of it being in
state $\left\vert  0_{e}1_{s}\right\rangle $.  For  our  two component
system we found these probabilities by solving  the TDSE with the ring
and field mode   parameters given above.  As  there is no  exchange of
energy between  the ring and the field   mode at point $B$,  it seemed
reasonable to assume that at this flux  bias the system would maintain
equal  probability  of  being in  each    of the   states  $\left\vert
  1_{e}0_{s}\right\rangle $ and $\left\vert 0_{e} 1_{s}\right\rangle $
at   times after $t_{0}$. The   results  of our  calculations of these
probabilities  are presented in   figure~\ref{fig:probAndEntangle}(a),
plotted against dimensionless time  $\omega_{e}t$.  We see from  these
computed solutions  that  our  assumption is essentially  correct,  at
least for the particular system example being studied.

Given  this prediction of the  behaviour of the  circuit system, it is
important to establish a measure of  the level of entanglement between
its two components.  In this  work  we adopt  the entropic measure  of
entanglement first introduced by Adami and Cerf~\cite{CerfA98}, i.e.
\[
I_{i}=S\left(  \rho\right)  -S\left(  \rho_{i}\right)
\]
where $S\left(  \rho\right) =Tr\left( \rho\ln\rho\right)  $ and $\rho$
is the density operator. If $I_{i}$ is  negative this implies that the
$i^{\mathrm{th}}$ component of  the system is  entangled with the rest
of the  system,  an inequality  that  holds  even  in  the presence of
dissipation.  This computed  quantity   for  our  system is  shown  in
figure~\ref{fig:probAndEntangle} (b)  as a function  of $\omega_{e}t$. 
As  we might suspect, $t_{0}$ is  also  the time when the entanglement
between the em  field and the SQUID ring  reaches a maximum.  From the
figure we see  that after approximately  a time $t_{0}$ both  the ring
and     field    mode probabilities   have    reached  values   of 0.5
(figure~\ref{fig:probAndEntangle}(a)) which  then remain constant with
time.  Similarly,  in~\ref{fig:probAndEntangle}(b) beyond  $t_{0}$ the
entanglement is strong (at 0.7) and also constant with time.

\begin{figure}[tb]
\begin{center}
\resizebox*{0.38\textwidth}{!}{\includegraphics{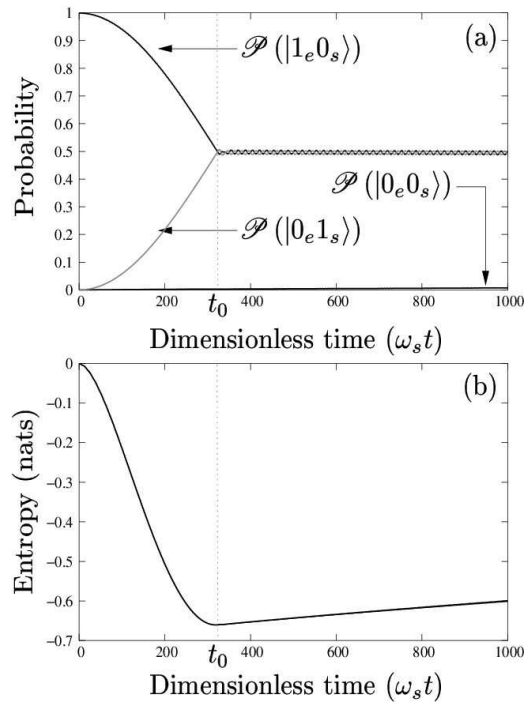}}
\caption{For comparison with figure~\ref{fig:probAndEntangle} (a)
probabilities of being in the state $\left\vert 1_{e}0_{s}\right\rangle $ or
$\left\vert 0_{e}1_{s}\right\rangle $ and (b) entanglement entropies $I_{s}$
and $I_{e}$ for the system evolving in a dissipative environment at point A
with the initial condition $\left\vert 1_{e}0_{s}\right\rangle $
.\vspace*{-0.3in}}
\label{fig:weakDissipation}
\end{center}
\end{figure}

In the above  discussion we have seen  from solutions of the TDSE that
we can use a  time dependent flux (\ref{fluxTimeFunction}), applied to
the  SQUID ring, to   make a maximally   entangled  state of  the form
(\ref{maxEntangledState}).  Nevertheless, for  this entanglement to be
of   use technologically it must   be  maintained over some reasonable
period of time in the  presence of dissipation.   In order to estimate
the effects of dissipation we adopt the  open systems approach usually
used in the field of quantum optics. In this approach the evolution of
the system   is found by  solving a  master equation for   the density
operator. The   master  equation  for  a  system subject    to thermal
dissipation is given by
\begin{align*}
\frac{\partial}{\partial t}\rho &  =\left[  H,\rho\right]  +\sum_{i=e,s}
\frac{\gamma_{i}}{2\hbar}\left(  M_{i}+1\right)  \left(  2a_{i}\rho
a_{i}^{\dagger}-a_{i}^{\dagger}a_{i}\rho-\rho a_{i}^{\dagger}a_{i}\right) \\
&  +\frac{\gamma_{i}}{2\hbar}M_{i}\left(  2a_{i}^{\dagger}\rho a_{i}
-a_{i}a_{i}^{\dagger}\rho-\rho a_{i}a_{i}^{\dagger}\right)
\end{align*}
where $\gamma_{i}$ is  the damping rate  of each system to the thermal
bath and the  mean  photon number  $M_{i}=2.746\times10^{-5}$  in each
component is related to the temperature $T_{b_{i}}=4.2$K and frequency
$\omega_{b_{i}}=\omega_{s}$  of each decohering bath via $M_{i}=\left[
  \exp\left( \hbar\omega_{b_{i}}/k_{B}T\right) -1\right] ^{-1}$.

\begin{figure}[t]
\begin{center}
\resizebox*{0.38\textwidth}{!}{\includegraphics{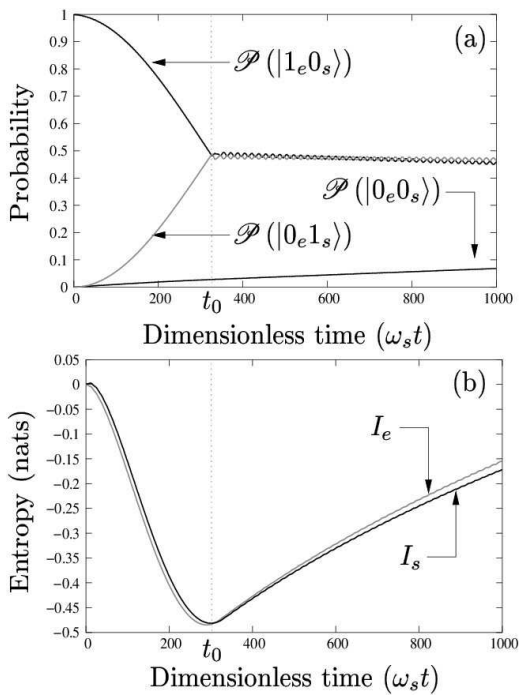}}
\caption{For comparison with figure~\ref{fig:probAndEntangle} (a)
probabilities of being in state $\left\vert 1_{e}0_{s}\right\rangle $ or
$\left\vert 0_{e}1_{s}\right\rangle $ and (b) entanglement entropies $I_{s}$
and $I_{e}$ in a strongly dissipative environment at point A with initial
condition $\left\vert 1_{e}0_{s}\right\rangle $.\vspace*{-0.3in}}
\label{fig:stronDissipation}
\end{center}
\end{figure}

The  effect  of dissipation   on  the probabilities  and  entanglement
entropies     of    our    chosen        system   are    shown      in
figures~\ref{fig:weakDissipation}  and~\ref{fig:stronDissipation}  for
the   cases    of weakly   and     strongly  dissipative environments,
respectively. Thus,  in figure~\ref{fig:weakDissipation} at a $\gamma$
value  of $1\times10^{-5} \omega_{s}$,  the effect on the probablities
is negligible over a few $t_{0}$ while the entanglement can be seen to
weaken only  gradually over the same  time period. For  this degree of
decoherence the corrsponding decoherence time is of the order of 20ns,
which is in line  with recent experimental results  on superconducting
weak link  circuits~\cite{Martinis2002}  (albeit at much   larger weak
link     capacitances   ($\approx10^{-12}$F) and  lower   temperatures
($\approx25$mK)). This suggests that at this  level of dissipation the
coherence of the coupled system could be  used effectively for quantum
coherent circuit operations.   It  is also apparent  from   the second
example of the effect of  a dissipative environment that substantially
greater  values  of $\gamma\left(  =1\times10^{-4}\omega_{s}\right)  $
i.e. stronger dissipation, still do not entirely suppress entanglement
over this  same time scale of a  few $t_{0}$. As can  be seen, at this
higher level of dissipation our system  starts to become mixed after a
sufficient  evolution  in  time. Concomitantly,    a time independent,
maximally entangled, state   is not maintained. Clearly,  this  result
suggests  a practical limit to  the level of dissipation (decoherence)
that  can  be tolerated if  quantum  coherent  operations   are to  be
performed successfully. With this in mind,  we note that these results
do not necessarily imply that dissipative effects will be significant,
although when  designing  real circuits environmental effects  must be
accounted for in its design.

In this paper  we have shown that, via  a time dependent external bias
flux,   a SQUID ring  can be  used to control  the  quantum state of a
coupled  em field  mode-SQUID  ring system,  including   the degree of
entanglement between  these two circuit components.  As a corollary of
this, we showed that  the level of this  entanglement could be set and
maintained, with only a slow degradation  in time, when the system was
coupled to a weakly dissipative external environment. With the current
interest  in using superconducting  weak   link circuits for   quantum
technologies, this  work provides some  guide to the time  scales over
which it may be possible to perform quantum coherent operations in the
presence of decohering environments.

\begin{acknowledgments}
We would like to thank Dr. T.P. Spiller for interesting discussions on
entanglement in quantum circuits.
\end{acknowledgments}

\vspace*{20pt}

\bibliographystyle{apsrev}
\bibliography{qcg}

\begin{thebibliography}{14}
\expandafter\ifx\csname natexlab\endcsname\relax\def\natexlab#1{#1}\fi
\expandafter\ifx\csname bibnamefont\endcsname\relax
  \def\bibnamefont#1{#1}\fi
\expandafter\ifx\csname bibfnamefont\endcsname\relax
  \def\bibfnamefont#1{#1}\fi
\expandafter\ifx\csname citenamefont\endcsname\relax
  \def\citenamefont#1{#1}\fi
\expandafter\ifx\csname url\endcsname\relax
  \def\url#1{\texttt{#1}}\fi
\expandafter\ifx\csname urlprefix\endcsname\relax\def\urlprefix{URL }\fi
\providecommand{\bibinfo}[2]{#2}
\providecommand{\eprint}[2][]{\url{#2}}

\bibitem[{\citenamefont{Friedman et~al.}(2000)\citenamefont{Friedman, Patel,
  Chen, Tolpygo, and Lukens}}]{FriedmanPCTL00}
\bibinfo{author}{\bibfnamefont{J.~R.} \bibnamefont{Friedman}},
  \bibinfo{author}{\bibfnamefont{V.}~\bibnamefont{Patel}},
  \bibinfo{author}{\bibfnamefont{W.}~\bibnamefont{Chen}},
  \bibinfo{author}{\bibfnamefont{S.~K.} \bibnamefont{Tolpygo}},
  \bibnamefont{and} \bibinfo{author}{\bibfnamefont{J.~E.}
  \bibnamefont{Lukens}}, \bibinfo{journal}{Nature}
  \textbf{\bibinfo{volume}{406}}, \bibinfo{pages}{43} (\bibinfo{year}{2000}).

\bibitem[{\citenamefont{van~der Wal et~al.}(2000)\citenamefont{van~der Wal, ter
  Haar, Wilhelm, Schouten, Harmans, Orlando, Lloyd, and
  Mooij}}]{vanderWalWSHOLM00}
\bibinfo{author}{\bibfnamefont{C.~H.} \bibnamefont{van~der Wal}},
  \bibinfo{author}{\bibfnamefont{A.~C.~J.} \bibnamefont{ter Haar}},
  \bibinfo{author}{\bibfnamefont{F.~K.} \bibnamefont{Wilhelm}},
  \bibinfo{author}{\bibfnamefont{R.~N.} \bibnamefont{Schouten}},
  \bibinfo{author}{\bibfnamefont{C.~J. P.~M.} \bibnamefont{Harmans}},
  \bibinfo{author}{\bibfnamefont{T.~P.} \bibnamefont{Orlando}},
  \bibinfo{author}{\bibfnamefont{S.}~\bibnamefont{Lloyd}}, \bibnamefont{and}
  \bibinfo{author}{\bibfnamefont{J.~E.} \bibnamefont{Mooij}},
  \bibinfo{journal}{Science} \textbf{\bibinfo{volume}{290}},
  \bibinfo{pages}{773} (\bibinfo{year}{2000}).

\bibitem[{\citenamefont{Rouse et~al.}(1995)\citenamefont{Rouse, Han, and
  Lukens}}]{RouseHL95}
\bibinfo{author}{\bibfnamefont{R.}~\bibnamefont{Rouse}},
  \bibinfo{author}{\bibfnamefont{S.~Y.} \bibnamefont{Han}}, \bibnamefont{and}
  \bibinfo{author}{\bibfnamefont{J.~E.} \bibnamefont{Lukens}},
  \bibinfo{journal}{Phys. Rev. Lett.} \textbf{\bibinfo{volume}{75}},
  \bibinfo{pages}{1614} (\bibinfo{year}{1995}).

\bibitem[{\citenamefont{Silvestrini et~al.}(2000)\citenamefont{Silvestrini,
  Ruggiero, Granata, and Esposito}}]{SilvestriniRGE00}
\bibinfo{author}{\bibfnamefont{P.}~\bibnamefont{Silvestrini}},
  \bibinfo{author}{\bibfnamefont{B.~B.} \bibnamefont{Ruggiero}},
  \bibinfo{author}{\bibfnamefont{C.}~\bibnamefont{Granata}}, \bibnamefont{and}
  \bibinfo{author}{\bibfnamefont{E.}~\bibnamefont{Esposito}},
  \bibinfo{journal}{Phys. Lett. A} \textbf{\bibinfo{volume}{267}},
  \bibinfo{pages}{45} (\bibinfo{year}{2000}).

\bibitem[{\citenamefont{Nakamura et~al.}(1999)\citenamefont{Nakamura, Pashkin,
  and Tsai}}]{NakamuraPT99}
\bibinfo{author}{\bibfnamefont{Y.}~\bibnamefont{Nakamura}},
  \bibinfo{author}{\bibfnamefont{Y.~A.} \bibnamefont{Pashkin}},
  \bibnamefont{and} \bibinfo{author}{\bibfnamefont{J.~S.} \bibnamefont{Tsai}},
  \bibinfo{journal}{Nature} \textbf{\bibinfo{volume}{398}},
  \bibinfo{pages}{786} (\bibinfo{year}{1999}).

\bibitem[{\citenamefont{Orlando et~al.}(1999)\citenamefont{Orlando, Mooij,
  Tian, van~der Wal, Levitov, Lloyd, and Mazo}}]{OrlandoMTvLLM99}
\bibinfo{author}{\bibfnamefont{T.~P.} \bibnamefont{Orlando}},
  \bibinfo{author}{\bibfnamefont{J.~E.} \bibnamefont{Mooij}},
  \bibinfo{author}{\bibfnamefont{L.}~\bibnamefont{Tian}},
  \bibinfo{author}{\bibfnamefont{C.~H.} \bibnamefont{van~der Wal}},
  \bibinfo{author}{\bibfnamefont{L.~S.} \bibnamefont{Levitov}},
  \bibinfo{author}{\bibfnamefont{S.}~\bibnamefont{Lloyd}}, \bibnamefont{and}
  \bibinfo{author}{\bibfnamefont{J.~J.} \bibnamefont{Mazo}},
  \bibinfo{journal}{Phys. Rev. B-Condens Matter} \textbf{\bibinfo{volume}{60}},
  \bibinfo{pages}{15398} (\bibinfo{year}{1999}).

\bibitem[{\citenamefont{Makhlin et~al.}(1999)\citenamefont{Makhlin, Schon, and
  Shnirman}}]{MakhlinSS99}
\bibinfo{author}{\bibfnamefont{Y.}~\bibnamefont{Makhlin}},
  \bibinfo{author}{\bibfnamefont{G.}~\bibnamefont{Schon}}, \bibnamefont{and}
  \bibinfo{author}{\bibfnamefont{A.}~\bibnamefont{Shnirman}},
  \bibinfo{journal}{Nature} \textbf{\bibinfo{volume}{398}},
  \bibinfo{pages}{305} (\bibinfo{year}{1999}).

\bibitem[{\citenamefont{Averin et~al.}(1990)\citenamefont{Averin, Nazarov, and
  Odintsov}}]{AverinNO90}
\bibinfo{author}{\bibfnamefont{D.~V.} \bibnamefont{Averin}},
  \bibinfo{author}{\bibfnamefont{Y.~V.} \bibnamefont{Nazarov}},
  \bibnamefont{and} \bibinfo{author}{\bibfnamefont{A.~A.}
  \bibnamefont{Odintsov}}, \bibinfo{journal}{Physica B}
  \textbf{\bibinfo{volume}{165}}, \bibinfo{pages}{945} (\bibinfo{year}{1990}).

\bibitem[{\citenamefont{Clark et~al.}(1998)\citenamefont{Clark, Diggins, Ralph,
  Everitt, Prance, Prance, Whiteman, Widom, and Srivastava}}]{ClarkDREPPWS98}
\bibinfo{author}{\bibfnamefont{T.~D.} \bibnamefont{Clark}},
  \bibinfo{author}{\bibfnamefont{J.}~\bibnamefont{Diggins}},
  \bibinfo{author}{\bibfnamefont{J.~F.} \bibnamefont{Ralph}},
  \bibinfo{author}{\bibfnamefont{M.}~\bibnamefont{Everitt}},
  \bibinfo{author}{\bibfnamefont{R.~J.} \bibnamefont{Prance}},
  \bibinfo{author}{\bibfnamefont{H.}~\bibnamefont{Prance}},
  \bibinfo{author}{\bibfnamefont{R.}~\bibnamefont{Whiteman}},
  \bibinfo{author}{\bibfnamefont{A.}~\bibnamefont{Widom}}, \bibnamefont{and}
  \bibinfo{author}{\bibfnamefont{Y.~N.} \bibnamefont{Srivastava}},
  \bibinfo{journal}{Annals Phys.} \textbf{\bibinfo{volume}{268}},
  \bibinfo{pages}{1} (\bibinfo{year}{1998}).

\bibitem[{\citenamefont{Everitt et~al.}(2001)\citenamefont{Everitt,
  T.~D.~Clark, Stiffell, Prance, Prance, Vourdas, and Ralph}}]{Everitt2001b}
\bibinfo{author}{\bibfnamefont{M.}~\bibnamefont{Everitt}},
  \bibinfo{author}{\bibfnamefont{T.}~\bibnamefont{T.~D.~Clark}},
  \bibinfo{author}{\bibfnamefont{P.}~\bibnamefont{Stiffell}},
  \bibinfo{author}{\bibfnamefont{H.}~\bibnamefont{Prance}},
  \bibinfo{author}{\bibfnamefont{R.}~\bibnamefont{Prance}},
  \bibinfo{author}{\bibfnamefont{A.}~\bibnamefont{Vourdas}}, \bibnamefont{and}
  \bibinfo{author}{\bibfnamefont{J.}~\bibnamefont{Ralph}}, \bibinfo{journal}{to
  be published in Phys Rev B} \textbf{\bibinfo{volume}{64}}
  (\bibinfo{year}{2001}).

\bibitem[{\citenamefont{Al-Saidi and Stroud}(2002)}]{Al-SaidiS02}
\bibinfo{author}{\bibfnamefont{W.~A.} \bibnamefont{Al-Saidi}} \bibnamefont{and}
  \bibinfo{author}{\bibfnamefont{D.}~\bibnamefont{Stroud}},
  \bibinfo{journal}{Phys. Rev. B} \textbf{\bibinfo{volume}{6501}},
  \bibinfo{pages}{art. no.} (\bibinfo{year}{2002}).

\bibitem[{\citenamefont{Likharev}(1986)}]{Likharev}
\bibinfo{author}{\bibfnamefont{K.~K.} \bibnamefont{Likharev}},
  \emph{\bibinfo{title}{Dynamics of Josephson Junctions and Circuits}}
  (\bibinfo{publisher}{Gordon and Breach}, \bibinfo{year}{1986}).

\bibitem[{\citenamefont{Cerf and Adami}(1998)}]{CerfA98}
\bibinfo{author}{\bibfnamefont{N.~J.} \bibnamefont{Cerf}} \bibnamefont{and}
  \bibinfo{author}{\bibfnamefont{C.}~\bibnamefont{Adami}},
  \bibinfo{journal}{Physica D} \textbf{\bibinfo{volume}{120}},
  \bibinfo{pages}{62} (\bibinfo{year}{1998}).

\bibitem[{\citenamefont{Martinis et~al.}(2002)\citenamefont{Martinis, Nam,
  Aumentado, and Urbina}}]{Martinis2002}
\bibinfo{author}{\bibfnamefont{J.~M.} \bibnamefont{Martinis}},
  \bibinfo{author}{\bibfnamefont{S.}~\bibnamefont{Nam}},
  \bibinfo{author}{\bibfnamefont{J.}~\bibnamefont{Aumentado}},
  \bibnamefont{and} \bibinfo{author}{\bibfnamefont{C.}~\bibnamefont{Urbina}},
  \bibinfo{journal}{Phys. Rev. Lett.} \textbf{\bibinfo{volume}{89}},
  \bibinfo{pages}{117901} (\bibinfo{year}{2002}).

\end{thebibliography}

\end{document}